%% file: MAIN.tex
\documentclass[10pt,twocolumn,letterpaper]{article} 

\usepackage{avss}
\usepackage{times}
\usepackage{epsfig}
\usepackage{graphicx}
\usepackage{amsmath}
\usepackage{amssymb}

\usepackage{algorithmic}
\usepackage[linesnumbered,ruled,vlined]{algorithm2e}
\SetKwInput{KwInput}{Input}                
\SetKwInput{KwOutput}{Output}              
\SetKw{Return}{return}

\usepackage{bm} 

\usepackage{dutchcal}

\let\oldnl\nl
\newcommand{\nonl}{\renewcommand{\nl}{\let\nl\oldnl}}

\usepackage{enumitem}
\setitemize{noitemsep,topsep=0pt,parsep=0pt,partopsep=0pt}

\usepackage{color, colortbl}
\definecolor{Gray}{gray}{0.9}
\definecolor{Yellow}{rgb}{1.0, 0.97, 0.7}

\usepackage{booktabs}

\usepackage{caption}
\usepackage{subcaption}

\usepackage{amsthm}
\theoremstyle{theoremstyle}

\usepackage{multirow}

\usepackage{color,soul} 

\avssfinalcopy 

\def\avssPaperID{92} 

\ifavssfinal\fi

\begin{document}

\title{Fast Private Location-based Information Retrieval over the Torus}



\author{Joon Soo Yoo$^1$, Mi Yeon Hong$^1$, Ji Won Heo$^2$, Kang Hoon Lee$^1$, Ji Won Yoon$^1$ \\
$^1$School of Cybersecurity, Korea University\\
$^2$Department of Physics, Korea University \\
{\tt\small \{sandiegojs, hachikohmy, hjw4, hoot55, jiwon\_yoon\}@korea.ac.kr}
}


\maketitle

\input{Section/00_abstract}

\input{Section/01_introduction}

\input{Section/02_background}
\input{Section/03_our_model}
\input{Section/04_proposed_method}

\input{Section/05_experiment}
\input{Section/06_result}

\input{Section/07_conclusion}

{\small
\bibliographystyle{ieee}
\bibliography{bib}
}

\end{document}

%% file: Section/00_abstract.tex
\begin{abstract}
   Location-based services offer immense utility, but also pose significant privacy risks. In response, we propose \textsf{LocPIR}, a novel framework using homomorphic encryption (HE), specifically the TFHE scheme, to preserve user location privacy when retrieving data from public clouds. Our system employs TFHE's expertise in non-polynomial evaluations, crucial for comparison operations. \textsf{LocPIR} showcases minimal client-server interaction, reduced memory overhead, and efficient throughput. Performance tests confirm its computational speed, making it a viable solution for practical scenarios, demonstrated via application to a COVID-19 alert model. Thus, \textsf{LocPIR} effectively addresses privacy concerns in location-based services, enabling secure data sharing from the public cloud.
   \let\thefootnote\relax\footnote{979-8-3503-7428-5/24/\$31.00 ©2024 IEEE}
\end{abstract}

%% file: Section/01_introduction.tex
\thispagestyle{empty}

\section{Introduction}

Location-based services offer individuals various advantages, including personalized features like navigation assistance and targeted recommendations. Nevertheless, these services also carry inherent risks, primarily centered around the potential compromise of individuals' privacy and the undesired consequences of data aggregation and behavioral analysis.

To address these challenges, homomorphic encryption (HE) emerges as a powerful cryptographic technique that facilitates computations on encrypted data. Notably, HE offers a distinct advantage over the traditional client-server model, known as two-party computation, wherein the client submits encrypted data for private computation that is subsequently evaluated without decryption. Consequently, the client obtains the computation result while safeguarding their sensitive information, as all operations are conducted within the encrypted domain.

In this paper, we introduce a novel framework, called \textsf{LocPIR}, which utilizes HE to hide user location information during the retrieval of location-associated data from the public cloud. The core component of our system is TFHE (Fast Fully Homomorphic Encryption over the Torus)~\cite{chillotti2020tfhe}, a prominent branch of HE schemes distinguished by its efficient gate bootstrapping.

Our motivation stems from recognizing the inherent advantage of the TFHE scheme in facilitating non-polynomial evaluations, particularly in \emph{comparison} operations. In contrast, other notable branches of HE schemes, such as the BGV-family~\cite{brakerski2014leveled}, exhibit lower performance in evaluation and thus rarely employ comparison operations. Consequently, some literature~\cite{an2021privacy, jain2021shelbrs} focusing on the construction of location-based PIR relies on the client's computational power or increased communication overhead to address the challenges posed by non-linear evaluations. 

\begin{figure}[htb]
    \centering
    \begin{subfigure}[t]{0.23\textwidth}
        \includegraphics[width=\textwidth]{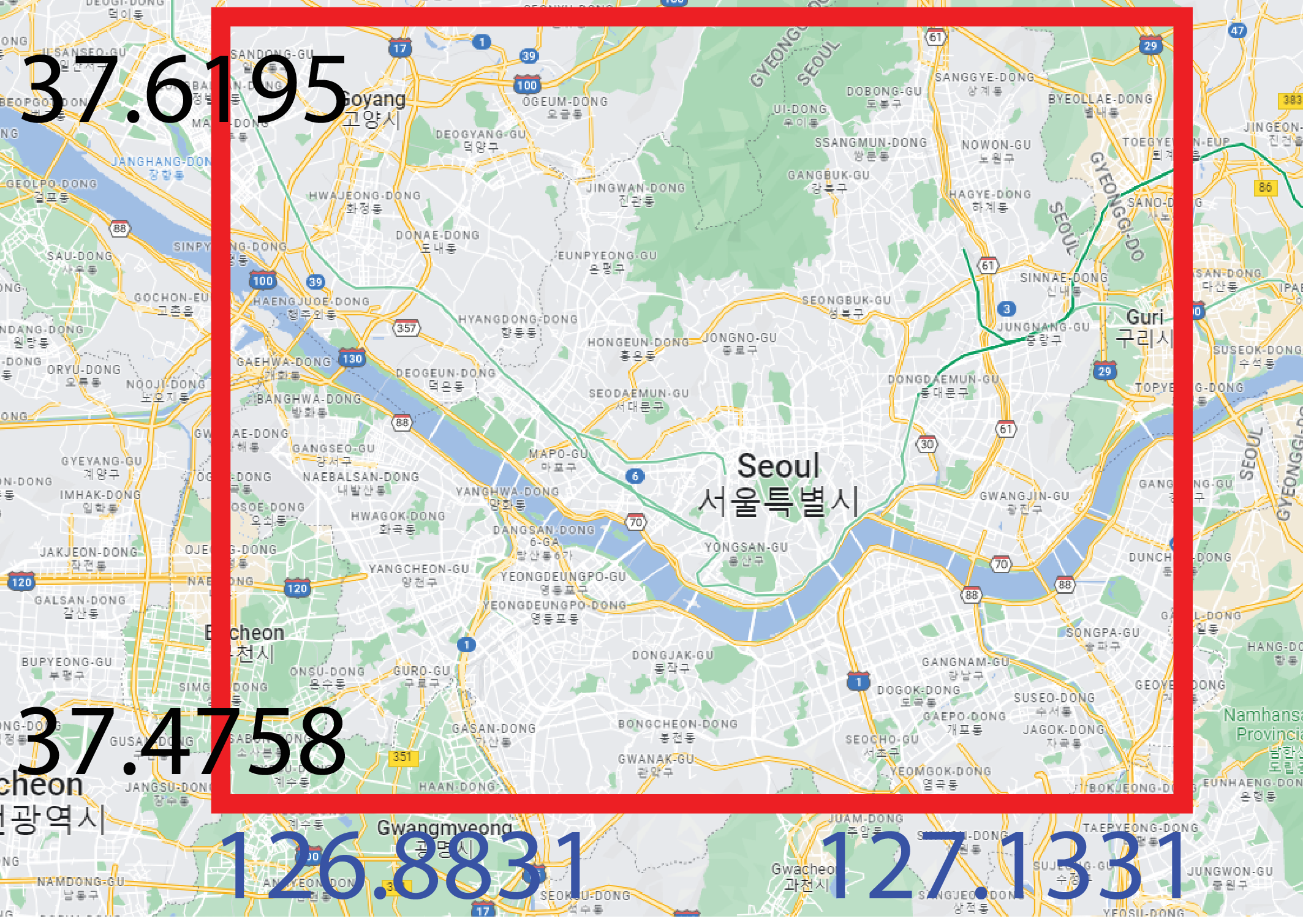}
        \caption{Seoul}
    \end{subfigure}
    \hfill
    \begin{subfigure}[t]{0.23\textwidth}
        \includegraphics[width=\textwidth]{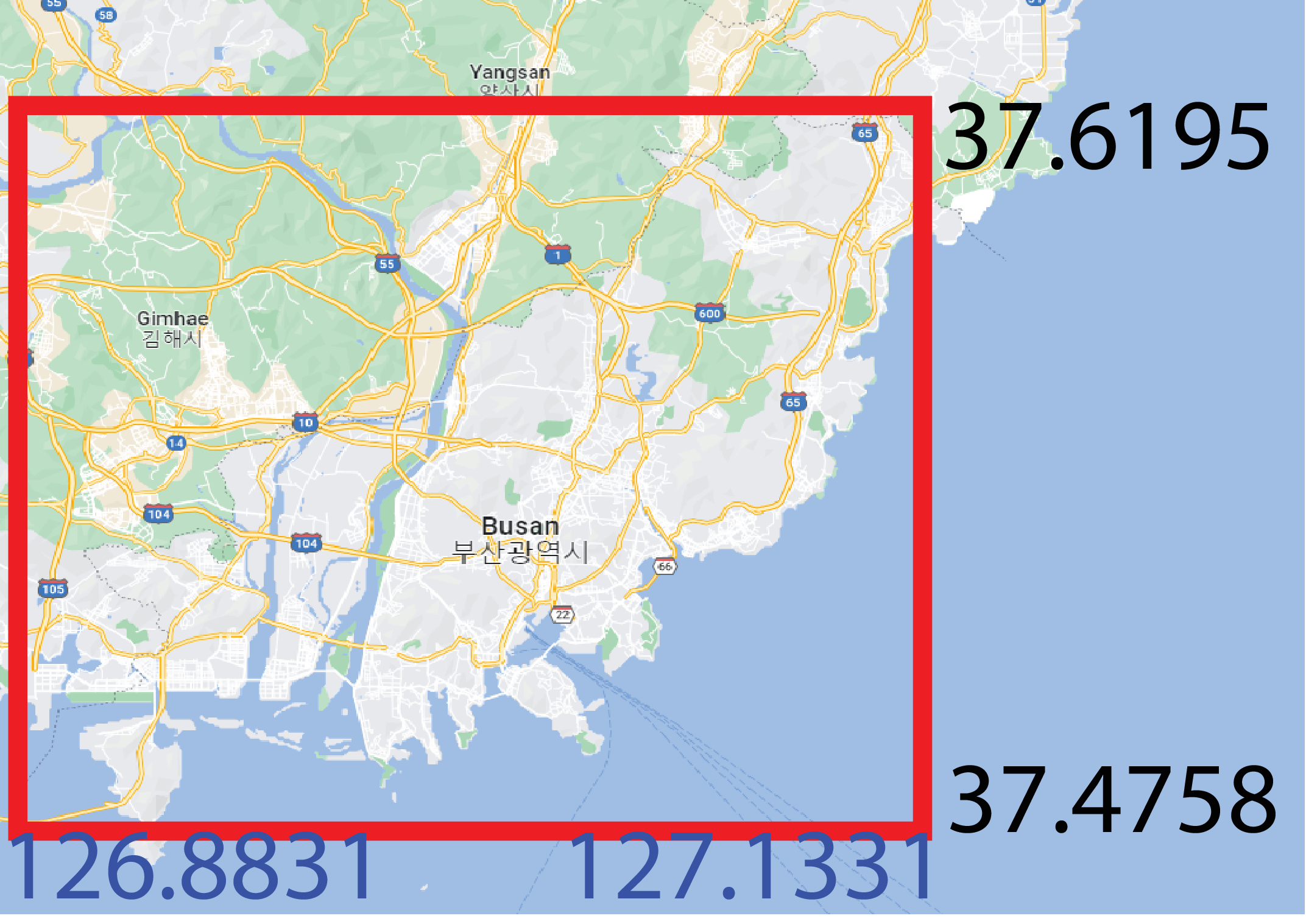}
        \caption{Busan}
    \end{subfigure}
    \hfill
    \caption{Bounding Box and Boundary Coordinates for Seoul and Busan.}
    \label{fig:korea_cities}
\end{figure}

In our study, we develop a homomorphic comparison circuit capable of processing two ciphertexts and generating a ciphertext that represents either greater than or less than relation, depending on the context. The primary objective of this construction is to establish a \emph{red bounding box}, as illustrated in Figure~\ref{fig:korea_cities}, that encompasses the geographical regions corresponding to the designated city for the encrypted service. Subsequently, upon receiving the user's encrypted GPS coordinates, the cloud can employ the \textsf{LocPIR} circuit to homomorphically compare these coordinates with the boundaries of the bounding box. The output of this computation allows the cloud to retrieve the \emph{encrypted service} associated with the relevant bounding box area. It is important to note that throughout this process, the server does not have access to the user's location information, as all computations are performed within the encrypted domain.

\subsection{Contribution}
\begin{itemize}[leftmargin=0.35cm]
    \item Our model is computationally efficient, enabling fast evaluation of user services in secure data sharing. It takes $5.12$ seconds for an $80$-bit security level and $5.67$ seconds for a $128$-bit level.
    \item The size of GPS ciphertexts in our framework is measured at $39.4$ KBytes, and the encrypted service result, specifically in our COVID-19 model, amounts to $22.9$ KBytes. These throughput values demonstrate the efficiency of our location-based PIR solution compared to other literature.
    \item Our framework is designed to minimize client-server interaction. At the preprocessing step, the client shares their public key, after which they only transmit encrypted GPS coordinates without further interaction. 
    \item We validate the practical feasibility of our model by applying it to a real-world scenario, specifically the COVID-19 alert model. 
\end{itemize}

\subsection{Related Work}

Jain et al.~\cite{jain2021shelbrs} propose a solution that involves two servers to determine the user's location. One server employs the Paillier cryptosystem for homomorphic addition, while the other server utilizes the Elgamal encryption scheme for multiplication. By leveraging partial homomorphic encryption, they aim to alleviate computational burdens. However, the assumption of non-collusion between the servers must be maintained. Additionally, their approach demonstrates a throughput of 5,000 data points in 4,600 seconds.

An et al.~\cite{an2021privacy} address location-based private information retrieval using a single-server setup, focusing on the COVID-19 alert model, which shares similarities with our problem setting. They employ the BFV scheme (BGV-family) to implement their model, requiring the user to compute the distance based on encrypted data received from the server. However, their approach has some drawbacks. Firstly, the computational overhead on the user's device consumes over 50\% of the CPU resources. Secondly, the total transfer size of ciphertext between the client and server amounts to 132.84 MBytes. Lastly, the total computation time required is approximately 399 seconds.

%% file: Section/02_background.tex
\section{Background}
\subsection{Notation} In the subsequent sections of this paper, we adopt the following notation: $\mathbb{B}$ represents the set $\{0, 1\}$, while $\mathbb{T}$ denotes the real Torus $\mathbb{R}/\mathbb{Z}$, which corresponds to the set of real numbers modulo $1$. The notation $a \stackrel{\$}{\leftarrow} \chi$ denotes uniform random sampling from the distribution $\chi$. The expression $\textsf{Enc}_{\mathfrak{K}}(m)$ refers to the encryption of the message $m$ using the secret key $\mathfrak{K}$. Throughout the paper, bold case letters, such as $\mathbf{a}$, are employed to denote vectors, while normal case letters are used to represent scalar values unless otherwise specified.

\subsection{Homomorphic Encryption}
Homomorphic encryption (HE) is a powerful encryption technique that enables the processing of encrypted data. The most promising solution against post-quantum attacks is LWE-based~\cite{regev2009lattices} (or RLWE-based~\cite{stehle2009efficient}) HE, which relies on the hardness of the Learning with Error (LWE) problem. The security of LWE-based encryption schemes heavily relies on introducing noise to the plaintext, making it computationally difficult to distinguish the encrypted form from random values (or LWE samples). HE schemes can be classified into two families: BGV-like~\cite{brakerski2012fully, fan2012somewhat, brakerski2014leveled}, and GSW-like~\cite{gentry2013homomorphic} schemes. 


\noindent\textbf{FHE over the Torus.} TFHE~\cite{chillotti2020tfhe} stands out as a leading scheme within the GSW-like family, primarily due to its highly efficient gate bootstrapping capability, with a single-core execution time of less than $13$ ms. The TFHE scheme encompasses three distinct types of ciphertexts that operate over the Torus $\mathbb{T}$. For the purpose of understanding our paper, we will specifically focus on the \textsf{TLWE} ciphertext type.

\begin{itemize}[leftmargin=0.35cm]
    \item \textsf{KeyGen}($\lambda$): Given security parameter $\lambda$, the key generation algorithm outputs a pair of keys: a secret key $\mathfrak{K} \stackrel{\$}{\leftarrow} \mathbb{B}$ and a public key set \textsf{PK}, which includes a bootstrapping key and a key switching key.  Additionally, the algorithm outputs \textsf{TLWE} parameters $n$ and $\sigma$, where both paramters are determined by the security paramter $\lambda$.
    \item \textsf{Encrypt}($m, \mathfrak{K}, n, \sigma$): Given a binary message $m \in \mathbb{B}$, it is encoded to a plaintext $\mu \in \mathbb{T}$ using the function $f: m \mapsto m/4 - 1/8$. Next, the algorithm generates a masking vector $\mathbf{a}$ by uniformly sampling from the set $U(\mathbb{T}^n)$. The plaintext message $\mu$ is then encrypted using the secret key $\mathfrak{K}$, resulting in a \textsf{TLWE} sample denoted as $\mathsf{ct} = \mathbf{a} \cdot \mathfrak{K} + \mu + e$. Here, $e$ is a noise term drawn from a Gaussian distribution $N(0, \sigma)$.
    \item \textsf{Decrypt}($\mathsf{ct}, \mathfrak{K}$): The decryption algorithm calculates the phase of the ciphertext $\mathsf{ct} = (\mathbf{a}, b)$ by $\varphi_{\mathfrak{K}}(\mathsf{ct}) = b - \mathbf{a} \cdot \mathfrak{K}$. The resulting phase $\mu + e$ is then rounded to the nearest plaintext message from the set $\{-1/8, 1/8\}$. Applying the inverse function $f^{-1}$ to the obtained plaintext message allows us to recover the original message bit $m \in \mathbb{B}$. Note that the error term $e$ must have a magnitude smaller than $1/16$ to guarantee the correctness of the algorithm.
    \item \textsf{EvalGate}($\mathsf{ct}_1, \mathsf{ct}_2$, \textsf{PK}): We focus on the evaluation of the \textsf{HomAND} gate as an illustrative example, noting that similar procedures apply to other basic gates (refer to [5] for details). The evaluation of the \textsf{HomAND} gate can be described as follows: 
    \begin{equation*}
        \textsf{HomAND}(\mathsf{ct}_1, \mathsf{ct}_2, \textsf{PK}) = \textsf{Bootstrap}((\mathbf{0}, -\frac{1}{8}) + \mathsf{ct}_1 + \mathsf{ct}_2).
    \end{equation*}
     Assuming that the magnitude of errors in both $\mathsf{ct}_1$ and $\mathsf{ct}_2$ is less than $1/16$, the aforementioned procedure correctly produces the \textsf{AND} result. We can verify by applying the phase function to $\mathsf{ct'} = (\mathbf{0}, -\frac{1}{8}) + \mathsf{ct}_1 + \mathsf{ct}_2$, resulting in: 
    \begin{equation*}
    \varphi_{\mathfrak{K}}(\mathsf{ct'}) = -\frac{1}{8} + \mu_1 + \mu_2 + e_1 + e_2.
    \end{equation*}
    If both $\mu_1$ and $\mu_2$ represent the encoding of $1$, indicating a result of \textsf{AND} as $1$, we can extract the plaintext message $\mu'$ from $\mathsf{ct'}$ when $\mu' > 0$. On the other hand, if $\mu'$ is less than $0$, we can determine that the result of \textsf{AND} is $0$.

\end{itemize}

%% file: Section/03_our_model.tex
\section{Our Model}

\begin{figure}[htb]
    \centering
    \includegraphics[width=0.47\textwidth]{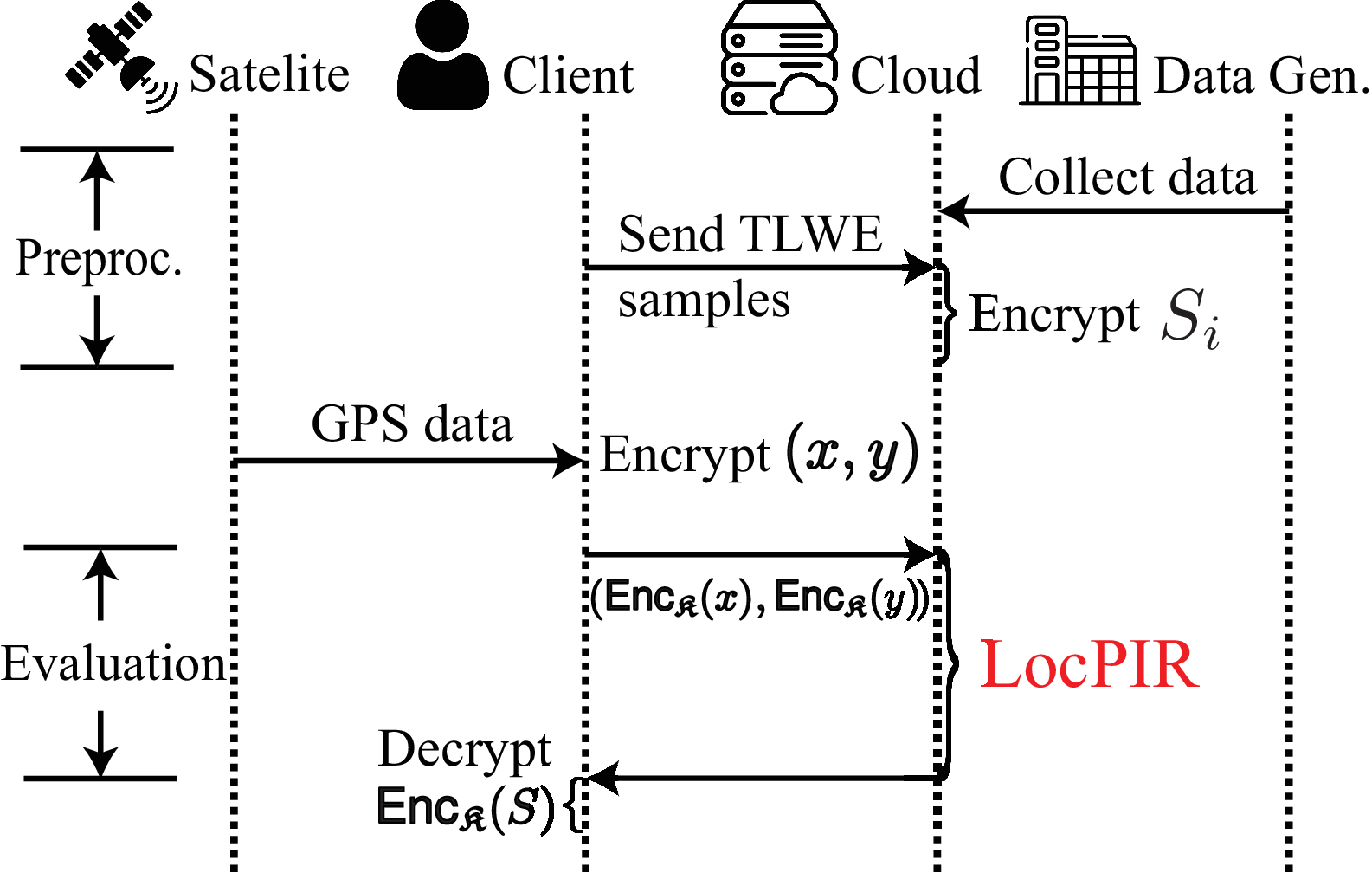}
    \caption{An Overview Timeline of Our Model.}
    \label{fig:our_overall_model}
\end{figure}

We present an overview of our model, which involves four distinct parties: a client, a cloud, a data generator, and a satellite (see Fig.~\ref{fig:our_overall_model}). We assume a \emph{semi-honest} cloud, which is curious about the client's information but adheres to the prescribed protocol. 

\noindent\textbf{Preprocessing.} The service provider collects data from a data generator, which, for instance, can be a COVID testing center responsible for aggregating the number of positive patients in a specific district. Upon receiving data from each data generating center, the cloud service provider generates service information, denoted as $S_i$, based on the received data and the corresponding geographical coordinates. For instance, the geographical coordinates of a city may span from $35.8$ to $38.2$ latitude and $125.2$ to $127.8$ longitude. The cloud service provider assigns a binary string, such as "immediate evacuation," to represent the generated service information $S_i$ for the client. Note that the length of the binary service string may vary; we introduce the parameter $m$ to denote the size of the service string.

The client transmits $N \times m$ freshly generated \textsf{TLWE} samples, represented as $\mathsf{ct}_{i, j}$, to the cloud encrypting the value of $0$ along side with the public key $\textsf{PK}$. These samples are subsequently utilized for encrypting the binary service string $S_i$ through the following steps: (1) encoding the individual bits of $S_i$ into a torus element, and (2) performing the operation $S_i[j] + \mathsf{ct}_{i, j}$. For instance, if the first bit of $S_i$ is $1$, it is encoded as $(0, 1/8)$ in the plaintext sample, and the fresh sample $(\textbf{a}_i, b_i)$ is added to it to transform it into a ciphertext (see Algorithm~\ref{alg:preprocess}). It should be noted that $\mathsf{ct}_{i, j}$ serves as a public key for the client, enabling the cloud to encrypt their binary services.

\begin{algorithm}[!htb]
\DontPrintSemicolon
  \KwInput{Trivial noiseless plaintext $S_i[j]$ and $\mathsf{ct}_{i, j} = (\mathbf{a}_{i, j}, b_{i, j})$ where $b_{i, j} = \mathbf{a}_{i, j} \cdot \mathfrak{K} + e$}
  \KwOutput{Encryption of $S_i[j]$ under secret key $\mathfrak{K}$}

  \For{$i=1,\dots,N$}
  {
    \For{$j=0,\dots,m-1$}
    {
        $S_i[j] \leftarrow S_i[j] + \mathsf{ct_{i,j}}$
    }
  }

  \Return{$S_i[j]$}
  
\caption{\textsf{Preprocess}($S_i[j], \mathsf{ct_{i,j}}$)}
\label{alg:preprocess}
\end{algorithm}

\noindent\textbf{Secure location-based computation.} The client obtains GPS data in the form of a tuple $(x,y)$ representing latitude $x$ and longitude $y$ from satellite sources. This data is encrypted using the client's secret key $\mathfrak{K}$ and transmitted to the cloud. The cloud receives the encrypted GPS data $(\textsf{Enc}_{\mathfrak{K}}(x), \textsf{Enc}_{\mathfrak{K}}(y))$ and performs an evaluation of the location-based service, denoted as \textsc{LocPIR}, utilizing the encrypted data and the client's public key $\textsf{PK}$. Upon completion of the evaluation, the cloud sends the encrypted result $S$ back to the client, which can only be decrypted by the client's secret key $\mathfrak{K}$.

%% file: Section/04_proposed_method.tex
\section{\textsc{LocPIR}}

\subsection{Encoding and Encryption} 

\noindent \textbf{Encoding.} We begin by explaining the encoding process for the input GPS data, denoted as $x$ and $y$. We adopt a fixed-point number representation methodology, as opposed to using a floating-point representation which requires more careful consideration. Given that latitude values range from $-90$ to $90$ degrees and longitude values range from $-180$ to $180$ degrees, we allocate a fixed $9$-bit representation to accommodate the integer part of the value. For the fractional part, we employ a flexible length, allowing for a customizable level of precision. By default, we assign a $7$-bit length for the fractional part, resulting in a total of $16$ bits for encoding the input GPS data $x$ and $y$. Note that the input length size is parameterized and denoted as $l$, indicating that our input has $l$-bit precision.

\noindent \textbf{Encryption.} In our encryption process, we adopt a bit-by-bit approach for the given encoded GPS data. Specifically, given a GPS data $x$, which has been encoded into a binary string of length $l$, we encrypt each individual bit of $x$ using the secret key $\mathfrak{K}$, which results in $l$ \textsf{TLWE} ciphertext. 

\subsection{Comparison Gate Design}
At the core of our framework lies a fundamental component---\textsf{HomCompS} circuit, representing a less-than comparison operation. This circuit accepts two ciphertexts, denoted as $\mathbf{c}_1$ and $\mathbf{c}_2$, which encrypt $x_1$ and $x_2$, each of length $l$ in binary format when encoded. The circuit produces a ciphertext of length $1$, wherein the encrypted value represents $1$ if $x_1 < x_2$, and $0$ otherwise. 

\begin{algorithm}[!htb]

\DontPrintSemicolon
  \KwInput{Ciphertext $\mathbf{c}_1$ and $\mathbf{c}_2$ of length $l$, encrypting $x_1$ and $x_2$ with secret key $\mathfrak{K}$ and \textsf{PK}}
  \KwOutput{$\textsf{Enc}_{\mathfrak{K}}$($1$) if $x_1 < x_2$, $\textsf{Enc}_{\mathfrak{K}}$($0$) otherwise}

  $t_0 \leftarrow \textsf{Enc}_{\mathfrak{K}}(0)$

  $\mathbf{c}_1[l-1] \leftarrow \textsf{HomNOT}(\mathbf{c}_1[l-1])$

  $\mathbf{c}_2[l-1] \leftarrow \textsf{HomNOT}(\mathbf{c}_2[l-1])$

  \For{$i=0,\dots,l-1$}
  {

    $t_1 \leftarrow \textsf{HomXNOR}(\mathbf{c}_1[i], \mathbf{c}_2[i])$

    $t_0 \leftarrow \textsf{HomMUX}(t_1, t_0, \mathbf{c}_2[i])$

  }

  \Return{$t_0$}
  
\caption{\textsf{HomCompS}($\mathbf{c}_1, \mathbf{c}_2$, \textsf{PK})}
\label{alg:homcomps}
\end{algorithm}

\noindent\textbf{Correctness.} The correctness of the circuit can be attributed to the iterative utilization of the \textsf{XNOR} gate and \textsf{MUX} gate within the loop defined in Algorithm~\ref{alg:homcomps}. The \textsf{XNOR} gate assigns $\textsf{Enc}_{\mathfrak{K}}(1)$ to $t_1$ when two bits are equal, while the \textsf{MUX} gate selects $t_0$ based on the previous result. Specifically, if the bits of $\mathbf{c}_1$ and $\mathbf{c}_2$ are equal, $t_0$ is chosen; otherwise, the corresponding bit of $\mathbf{c}_2$ is selected.

Considering the scenario where $x_1$ and $x_2$ are equal, the \textsf{XNOR} gate yields $t_0$ as the output, which is then propagated through the \textsf{MUX} gate, resulting in the selection of $t_0$ once again. Consequently, the initial value of $t_0$ (i.e., $\textsf{Enc}_{\mathfrak{K}}(0)$) is successfully propagated to the final output of the circuit, satisfying the desired outcome.

In the scenario where $\mathbf{c}_1$ and $\mathbf{c}_2$ differ, the focus is on the most significant differing position (MSDP), which represents the highest bit position where the corresponding bits of $\mathbf{c}_1$ and $\mathbf{c}_2$ differ. At the MSDP, the \textsf{MUX} gate selects the corresponding bit of $\mathbf{c}_2$ for $t_0$. Consequently, if $x_1 > x_2$, the desired outcome of $\textsf{Enc}_{\mathfrak{K}}(0)$ is achieved, indicating that $x_1$ is not less than $x_2$. A similar process applies to the opposite case.

To address potential issues, bit-flipping is introduced in line $2$ and $3$ of Algorithm~\ref{alg:homcomps}. In situations where $x_1$ and $x_2$ are equal, the iterative process passes $t_0$ from the MSDP bit to the output, ensuring correctness of the circuit. However, when $x_1$ and $x_2$ differ, the MSDP corresponds to the most significant bit (MSB), and the \textsf{MUX} gate selects the MSB of $\mathbf{c}_2$. This selection may yield an opposite result ($\textsf{Enc}_{\mathfrak{K}}(1)$) instead of the desired $\textsf{Enc}_{\mathfrak{K}}(0)$ when $x_1 > x_2$. Thus, the MSB needs to be initially reversed. Likewise, \textsf{HomCompLE} can be designed by negating the result from \textsf{HomCompS}, as it yields the inverse outcome.





  


\subsection{LocPIR Circuit}

\begin{algorithm}[!htb]
\DontPrintSemicolon
  \KwInput{Client's encrypted GPS coordinate $(\textsf{Enc}_{\mathfrak{K}}(x), (\textsf{Enc}_{\mathfrak{K}}(y))$ and \textsf{PK}}
  \KwOutput{Requested service $\textsf{Enc}_{\mathfrak{K}}(S)$}

  $S \leftarrow [\textsf{Enc}_{\mathfrak{K}}(0), \cdots, \textsf{Enc}_{\mathfrak{K}}(0)]$ \tcp{size $m$}

  \For{$i=1,\dots,N$}
  {
    \tcc{compare all intervals}
    $x_l \leftarrow \textsf{HomCompLE}(x_{i, 1}, \textsf{Enc}_{\mathfrak{K}}(x), \textsf{PK})$

    $x_r \leftarrow \textsf{HomCompS}(\textsf{Enc}_{\mathfrak{K}}(x), x_{i, 2}, \textsf{PK})$
    
    $y_l \leftarrow \textsf{HomCompLE}(y_{i, 1}, \textsf{Enc}_{\mathfrak{K}}(y), \textsf{PK})$

    $y_r \leftarrow \textsf{HomCompS}(\textsf{Enc}_{\mathfrak{K}}(y), y_{i, 2}, \textsf{PK})$

    \tcc{validation \#1}
    $v_1 \leftarrow \textsf{HomAND}(x_l, x_r, \textsf{PK})$

    $v_2 \leftarrow \textsf{HomAND}(y_l, y_r, \textsf{PK})$

    \tcc{validation \#2}
    $f \leftarrow \textsf{HomAND}(v_1, v_2, \textsf{PK})$

    \tcc{service validation}
    $S_i \leftarrow \textsf{bitwiseAND}(f, S_i, \textsf{PK})$

    \tcc{addition of all services}
    $S \leftarrow \textsf{HomSum}(S, S_i, \textsf{PK})$
  }

  \Return{$S$}
  
\caption{\textsf{LocPIR}($\textsf{Enc}_{\mathfrak{K}}(x), \textsf{Enc}_{\mathfrak{K}}(y)$, \textsf{PK})}
    \label{alg:locpir}
\end{algorithm}

\noindent\textbf{Comparison with bounding boxes.} The \textsf{LocPIR} circuit takes the client's encrypted GPS coordinates $(\textsf{Enc}_{\mathfrak{K}}(x), \textsf{Enc}_{\mathfrak{K}}(y))$, along with their public key \textsf{PK}, as inputs. The circuit then produces the requested service of the client, represented as $\textsf{Enc}_{\mathfrak{K}}(S)$, according to the algorithm outlined in Algorithm~\ref{alg:locpir}. The underlying concept behind this circuit is to compare the GPS coordinates with the bounding box coordinates $(x_{i, 1}, x_{i, 2}, y_{i, 1}, y_{i, 2})$, where each coordinate defines the latitude and longitude range of the $i-$th region. If the values of $x$ and $y$ fall within the boundary of the bounding box, the comparison result, as shown in line 3-6 of Algorithm~\ref{alg:locpir} will output $\textsf{Enc}_{\mathfrak{K}}(1)$.

\noindent\textbf{Validation step.} The obtained comparison result from line 3-6 allows us to verify whether the given GPS coordinate falls within the specified bounding box. This validation step occurs in line 7-10 of Algorithm~\ref{alg:locpir}. Utilizing the property of the homomorphic \textsf{AND} gate (\textsf{HomAND}), which outputs $\textsf{Enc}_{\mathfrak{K}}(1)$ only when all encrypted values are 1, we can determine the fulfillment of the bounding box condition. Specifically, if the GPS coordinates fall within the bounding box, we have $x_l = x_r = y_l = y_r = \textsf{Enc}_{\mathfrak{K}}(1)$. By applying \textsf{HomAND} gates, we can obtain $\textsf{Enc}_{\mathfrak{K}}(1)$. 

\begin{algorithm}[!htb]
\DontPrintSemicolon
  \KwInput{1-bit encryption $\mathsf{ct}$, $\mathbf{c}$ encrypting $m$-bit binary $x$, and \textsf{PK}}
  \KwOutput{(size $m$) $\mathbf{c}$ encrypting $x$ or $0$}

  \For{$i=0,\dots,m-1$}
  {
    $\mathbf{c}[i] \leftarrow \textsf{HomAND}(\mathsf{ct}, \mathbf{c}[i], \textsf{PK})$
  }
  
  \Return{$\mathbf{c}$}
  
\caption{\textsf{bitwiseAND}($\mathsf{ct}, \mathbf{c}$, \textsf{PK})}
\label{alg:bitAND}
\end{algorithm}

\noindent\textbf{Sevice validation.} Subsequently, we can perform a bitwise \textsf{HomAND} operation, denoted by \textsf{bitwiseAND} (see Algorithm~\ref{alg:bitAND}), on the previous validation result $f$ with $S_i[j]$ of length $m$ to obtain $S_i$, which represents the encrypted service $S_i$ associated with the $i$-th location or bounding box. If the validation step yields a result of $\textsf{Enc}_{\mathfrak{K}}(1)$ for $f$, it indicates that the GPS coordinates $x$ and $y$ fall within the $i$-th boundary. Consequently, the $i$-th encrypted service $S_i$ will be preserved in the \textsf{bitwiseAND} operation. 


\begin{figure}[!htb]
    \centering
    \includegraphics[width=0.47\textwidth]{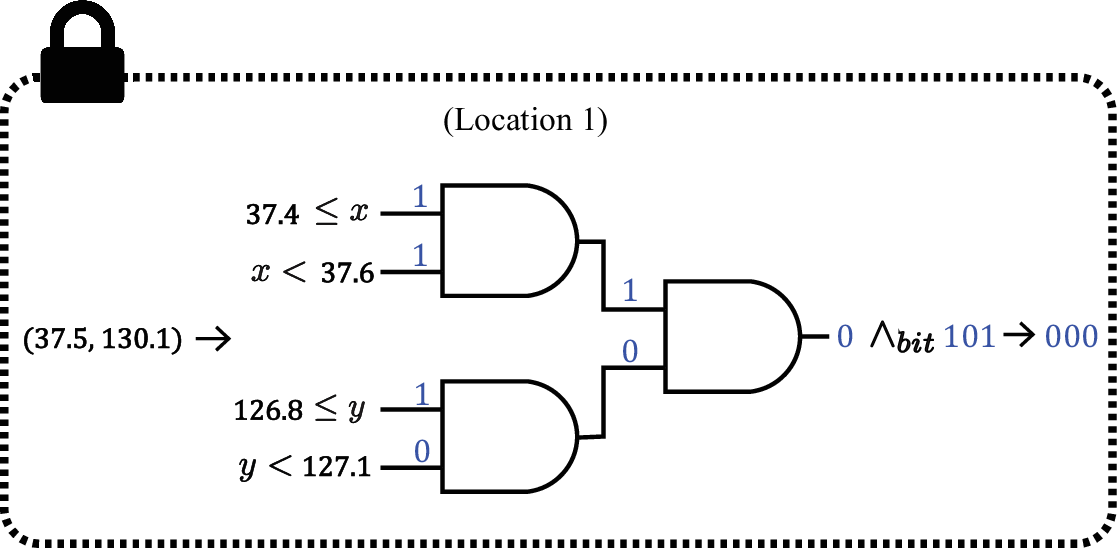}
    \caption{Diagram Illustrating Comparison Gate Evaluation and Validation Step.}
    \label{fig:easy_loc}
\end{figure}

\noindent\textbf{Illustrative example.} Consider the following scenario, illustrated in Fig.~\ref{fig:easy_loc}, where $x=37.5$ and $y=130.1$ in the plain domain. Initially, we compare $x$ and $y$ with $N$ location boundaries represented by $x_{i, 1}, x_{i, 2}, y_{i, 1}, y_{i, 2}$. For instance, in the case of location 1, the latitude boundary is defined as $[37.4, 37.6)$ and the longitude boundary as $[126.8, 127.1)$. The comparison results output $1$ for $x$ since it falls within the latitude boundary. Conversely, $y$ does not satisfy the boundary condition $y < 127.1$, resulting in a comparison outcome of $0$. Consequently, the bitwise \textsf{AND} operation applied to these outcomes yields $0$ as the final validation result ($f=0$). Thus, performing a bitwise \textsf{AND} of $f$ with the corresponding service $S_1=101$ in binary form yields $000$. In summary, if the input GPS coordinates fail to satisfy the bounding box condition, the resulting $S_i$ is represented as $000$ in binary. Conversely, if the GPS coordinates satisfy the bounding condition, the validation result $f$ will be $1$, preserving the original binary representation of $S_i$.

\subsection{Services Summation: \textsf{HomAddXOR}}
In the final step, we combine the results of $N$ services $S_i$. Under the assumption of disjointly distributed bounding boxes, there exists only one $S_i$ (or none) that corresponds to the encryption of its binary representation, while the remaining $S_i$ values are encryption of $0$. This distinction arises due to the unique matching of the GPS coordinate with a specific bounding box condition. Consequently, we can sum the $N$ services $S_i$, resulting in the service associated with the bounding box in which the coordinates $x$ and $y$ reside.

\noindent \textbf{XOR Addition.} An appealing approach is to employ \textsf{HomXOR} gates to perform the addition of the corresponding bits in $S_i$. For each bit position $j \in [0, m)$, we can compute the operation $S_1[j] \oplus \dots \oplus S_N[j]$ (see Algorithm~\ref{alg:homxor}). 



\begin{algorithm}[!htb]
\DontPrintSemicolon
  \KwInput{$\{S_i\}_{i=1}^N$, where at most one of $S_i$'s is not an encryption of $0$ and \textsf{PK}}
  \KwOutput{$S$ (sum of $S_i$'s)}

  $S \leftarrow [\textsf{Enc}_{\mathfrak{K}}(0), \cdots, \textsf{Enc}_{\mathfrak{K}}(0)]$ \tcp{size $m$}
  
  \For{$i=1,\dots,N$}
  {
    \For{$j=0,\dots,m-1$}
    {
        $S_i[j] \leftarrow \textsf{HomXOR}(S[j], S_i[j], \textsf{PK})$
    }
  }
  
  \Return{$S$}
  
\caption{\textsf{HomAddXOR}($\{S_i\}_{i=1}^N$, \textsf{PK})}
\label{alg:homxor}
\end{algorithm}

%% file: Section/05_experiment.tex
\section{Experiment}

\noindent \textbf{Environment Setting.} The experiments were performed on a computer with the following specifications: an AMD Ryzen 5 3500X 6-Core Processor and 64GB of RAM. The operating system used was Ubuntu 22.04 LTS. We implemented our model using the TFHE library version 1.1.

\begin{table}[htb!]
\centering
\scalebox{0.8}{
\begin{tabular}{cccccc}
\toprule
City & Lat1 & Lat2 & Long1 & Long2 & Service \\
\midrule
Seoul & 37.4758 & 37.6195 & 126.8831 & 127.1331 & 427 \\

\rowcolor{Gray}
Busan & 35.1692 & 35.2199 & 128.8821 & 129.2104 & 33 \\

Daegu & 35.7467 & 35.9743 & 128.4280 & 128.6909 & 61 \\

\rowcolor{Gray}
Incheon & 37.4127 & 37.5941 & 126.5876 & 126.7894 & 74 \\

Gwangju & 35.0732 & 35.2361 & 126.7050 & 126.9533 & 5 \\

\rowcolor{Gray}
Daejeon & 36.2615 & 36.4508 & 127.2998 & 127.5076 & 13 \\

Ulsan & 35.3710 & 36.6745 & 129.0731 & 129.3644 & 9 \\

\rowcolor{Gray}
Sejong & 36.5018 & 36.6970 & 127.1430 & 127.4080 & 6 \\

Jeju & 33.2948 & 33.5149 & 126.1874 & 126.8493 & 6 \\
\bottomrule
\end{tabular}}
\caption{COVID-19 Dataset for Nine Major Cities: Bounding Box Coordinates and Patient Incidence (Services).}
\label{tab:covid_dataset}
\end{table}

\noindent \textbf{Dataset.} We utilized the publicly available dataset provided by the Korea Disease Control and Prevention Agency (KDCA), which contains the daily records of COVID-19 patient incidence for the nine major cities in Korea. The dataset used in this study specifically focuses on October 26th, 2021. To define the geographical boundaries of each major city, we obtained the coordinates of the bounding box region using Google Maps (see Fig.~\ref{fig:korea_cities}).  Subsequently, we associated the COVID-19 patient incidence data for each city with the corresponding "Service" information. 


\noindent\textbf{TFHE Parameters.} Our model utilized standard 80-bit ($\lambda_{80}$) and 128-bit ($\lambda_{128}$) security levels. The selection of parameters $n$ and $\sigma$, which determine the security level of the LWE scheme, was chosen based on the lattice estimator~\cite{albrecht2015concrete}. Specifically, for the 80-bit security level, we set $n=540$ and $\sigma=2^{-20.2}$. Similarly, for the 120-bit security level, we chose $n=630$ and $\sigma=2^{-13.8}$.

\noindent \textbf{COVID-19 Model Outline.} The scenario for the COVID-19 alert model aligns closely with our previous proposal (See Fig.~\ref{fig:our_overall_model}). Based on the aggregated result of COVID-19 incidences, the cloud system evaluates \textsf{LocPIR} to provide the user with the encrypted service result, which can only be decrypted with the user's secret key.

%% file: Section/06_result.tex
\section{Result}

\subsection{Privacy-preserving COVID-19 Alert Model}

\begin{table*}[htb!]
\centering
\begin{tabular}{cccccccc}
\toprule
\multirow{2}{*}{Precision} & \multirow{2}{*}{Parameter Set} & \multirow{2}{*}{Encrypt} & \multicolumn{4}{c}{\textsf{LocPIR}} & \multirow{2}{*}{Decrypt} \\
\cline{4-7}
& & & Comparison & Validation & \textsf{HomAddXOR} & \textbf{Total Time} & \\
\midrule
\multirow{2}{*}{$l=13$} & $\lambda_{80}$ & $0.34 \, \text{ms}$ & $3.92 \, \text{s}$ & $0.32 \, \text{s}$ & $0.12 \, \text{s}$ & $\mathbf{4.36\, \text{s}}$ & $6.27 \, \text{µs}$ \\
 & $\lambda_{128}$ & $0.38 \, \text{ms}$ & $4.40\, \text{s}$ & $0.33\, \text{s}$ & $0.15 \, \text{s}$ & $\mathbf{4.88\, \text{s}}$ & $5.55 \, \text{µs}$ \\
\midrule
\multirow{2}{*}{$l=16$} & $\lambda_{80}$ & $0.41 \, \text{ms}$ & $4.69\, \text{s}$ & $0.31 \, \text{s}$ & $0.12 \, \text{s}$ & $\mathbf{5.12\, \text{s}}$ & $5.57 \, \text{µs}$ \\
 & $\lambda_{128}$ & $0.48 \, \text{ms}$ & $5.21\, \text{s}$ & $0.33\, \text{s}$ & $0.13 \, \text{s}$ & $\mathbf{5.67\, \text{s}}$ & $4.97 \, \text{µs}$ \\
\bottomrule
\end{tabular}
\caption{Execution Time of \textsf{LocPIR} COVID-19 Alert Model: Fixed Service Length and Major Cities Number ($m, N =9$).}
\label{tab:performance}
\end{table*}


\noindent\textbf{Time Performance.} The evaluation (\textsf{LocPIR}) process took a total of $4.36$ seconds for the 80-bit security level and $4.88$ seconds for the 120-bit security level when the input length was $13$. Additionally, for an input length of $16$, the evaluation required $5.12$ seconds for the 80-bit security level and $5.67$ seconds for the 120-bit security level. It is worth noting that the encryption and decryption stages each took less than a millisecond, regardless of the security parameters.

The comparison circuit evaluations, which involved encrypted GPS coordinates with $N$ bounding boxes, accounted for the majority of the execution time in the evaluation step. Specifically, for $l=13$ and $\lambda_{80}$, the comparison circuit evaluations took $3.92$ seconds. In comparison, the Validation step required $0.32$ seconds, and the \textsf{HomAddXOR} step took $0.12$ seconds. This highlights that the comparison circuit evaluations constituted approximately $90$ percent of the total evaluation time.

\noindent\textbf{Memory Size.} In our parameter setting, a \textsf{TLWE} sample for $\lambda_{80}$ occupies approximately $2.11$ KBytes of memory. As a result, during the preprocessing stage, the user is required to transmit $N \times m = 81$ \textsf{TLWE} samples for the encryption of services to the public cloud server, totaling around $171$ KBytes. Subsequently, the user transmits two encrypted GPS coordinates, which are approximately $27.4$ KBytes and $33.8$ KBytes in size for lengths $13$ and $16$, respectively. For $\lambda_{128}$, the user transmits approximately $199$ KBytes of $81$ \textsf{TLWE} samples, along with approximately $32$ KBytes and $39.4$ KBytes for size $13$ and $16$ GPS coordinates, respectively. 

\subsection{General Circuit Analysis.}
\noindent\textbf{Time Complexity Anlaysis.} We emphasize that the primary bottleneck in the \textsf{LocPIR} algorithm stems from the evaluation of comparison circuit, which involves $l$ \textsf{XNOR} and \textsf{MUX} gates, resulting in $3l$ evaluations of homomorphic gates (with \textsf{MUX} gates taking twice the computational time compared to other gates). In summary, the number of evaluations for the comparison circuits is in the order of $O(Nl)$. The validation step necessitates $O(N+m)$ evaluations. Lastly, summing up all the services with \textsf{HomAddXOR} requires $O(mN)$ evaluations. Thus, the total complexity of our model can be expressed as $O(N(m+l))$.

\begin{figure}[htb]
    \centering
    \begin{subfigure}[t]{0.23\textwidth}
        \includegraphics[width=\textwidth]{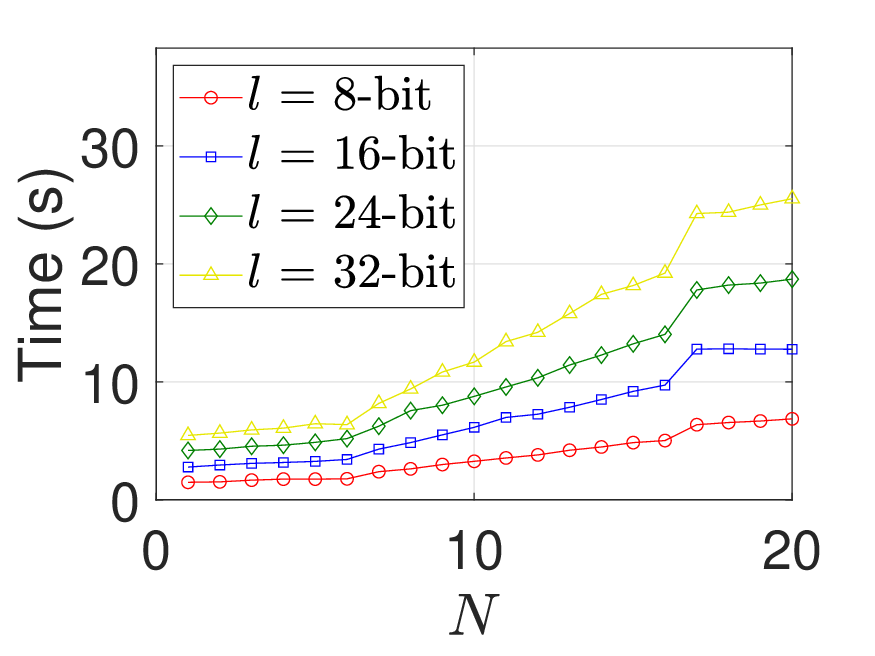}
        \caption{Time w.r.t $(N, l)$}
    \end{subfigure}
    \hfill
    \begin{subfigure}[t]{0.23\textwidth}
        \includegraphics[width=\textwidth]{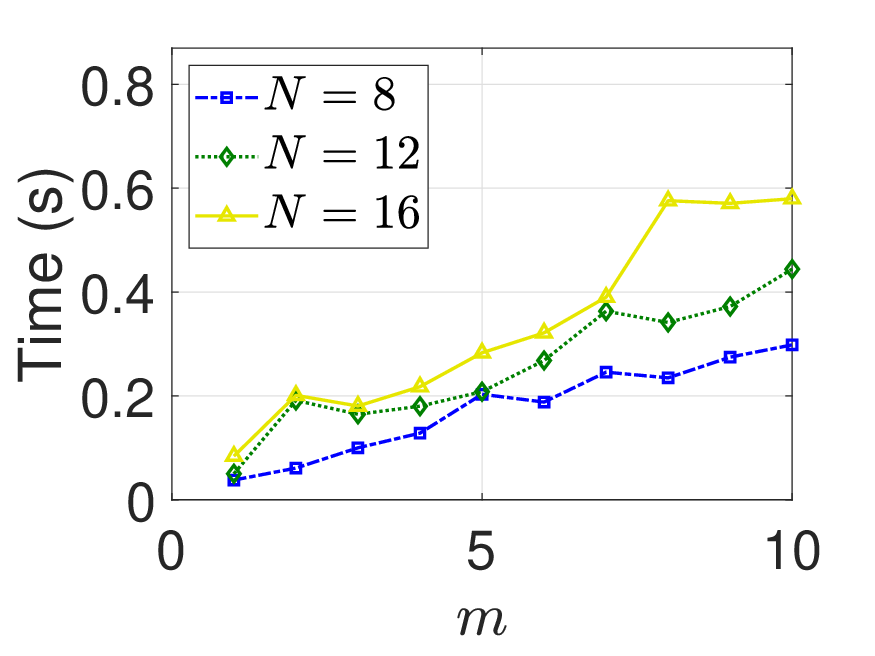}
        \caption{Time w.r.t $(m, N)$}
    \end{subfigure}
    \hfill
    \caption{\textsf{LocPIR}'s General Execution Time with Key Parameters: Number of Bounding Boxes $N$, Input Length $l$, and Service Length $m$.}
    \label{fig:N_l_m}
\end{figure}

\begin{figure}[!htb]
    \centering
    \includegraphics[width=0.47\textwidth]{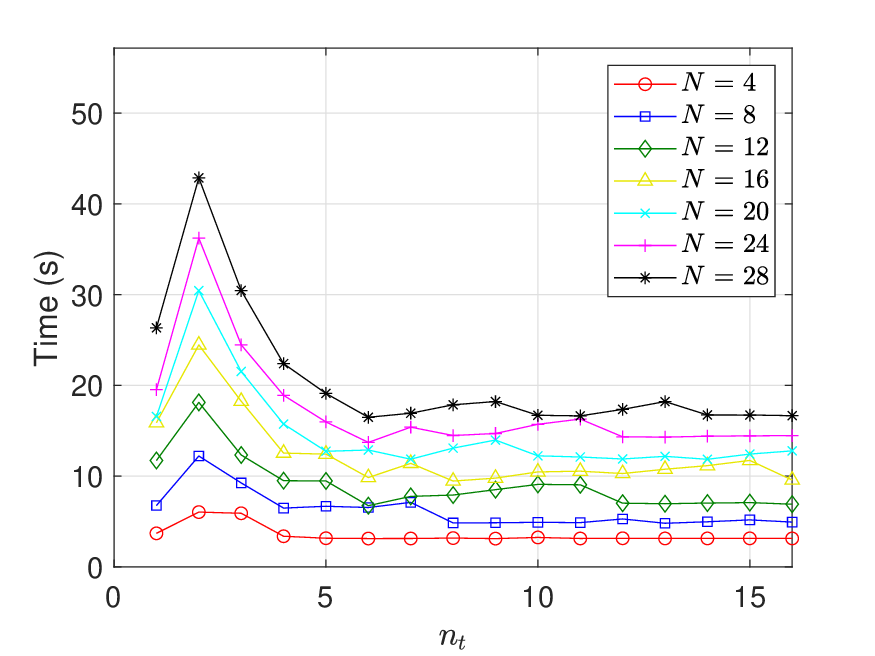}
    \caption{Execution Time w.r.t Thread Count $n_t$ for Varying Numbers of Bounding Boxes $N=4, 8, 12, 16, 20, 24, 28$.}
    \label{fig:by_threads}
\end{figure}

\noindent\textbf{Timing Analysis of Parameters $(N, l, m)$.} 
First, we conducted experiments on our model by varying the number of bounding boxes ($N$) and the input GPS length ($l$) (refer to Fig.~\ref{fig:N_l_m}.(a)). The service length was fixed at $m=9$, and full threading with $n_t = 6$ was utilized. It is observed that the slope of the graphs changes at $N=6$, where the magnitude of the slope is affected by the limited number of threads. Further validation of this claim can be seen in Fig.~\ref{fig:by_threads}.

In addition, we can observe a linear increase in execution time with respect to $l$. As an example of concrete experimental results, when $N=10$, the evaluation times for input lengths $l=8, 16, 24, 32$ bits are $3.256$, $6.139$, $8.764$, and $11.670$, respectively. These times correspond to $1.88\times$, $2.69\times$, and $3.58\times$ of the first evaluation time. 

Next, we examined the performance of our model by varying the service length $m$ and the number of bounding boxes $N$ while keeping the input length $l$ fixed at $16$ (see Fig.~\ref{fig:N_l_m}.(b)). The time performance demonstrated a linear relationship with the service length $m$, which is in line with our estimated time complexity.

\noindent\textbf{Highly Parallel Structure.} The structure of \textsf{LocPIR} is \emph{significantly parallelizable} (see Fig.~\ref{fig:by_threads}). The figure shows that an increase in thread count leads to a drastic improvement in time performance, specifically within the range of $2$ to $6$. In conclusion, a sufficiently large number of parallel computations can drastically enhance the performance of \textsf{LocPIR}. 


%% file: Section/07_conclusion.tex
\section{Conclusion}
This study introduces a location-based private information retrieval framework capable of concealing location data from a public cloud during information retrieval pertaining to GPS coordinates. It has been demonstrated that our system can be re-engineered homomorphically using the TFHE scheme, noted for its proficiency in executing non-linear polynomial evaluations such as comparisons. Furthermore, the applicability of our system within a COVID-19 alert model conforming to standard security levels has been validated. Given the highly parallel structure of our circuit, future research efforts will concentrate on exploring and enhancing these design improvements.

\vspace{0.3cm}

\noindent\textbf{Acknowledgments} This work was supported in part by the MSIT (Ministry of Science and ICT), Korea, under the ITRC (Information Technology Research Center) support program (IITP-2024-RS-2022-00164800) supervised by the IITP (Institute for Information \& Communications Technology Planning \& Evaluation), and in part by the grant-in-aid of LG Electronics.